# UV – laser modification and selective ion-beam etching of amorphous    vanadium pentoxide thin films

**Alexander Cheremisin**[*, 1], **Vadim Putrolaynen**[1], **Andrey Velichko**[**, 1], **Alexander Pergament**[1],

**Nikolay Kuldin**[1] **and Alexander Grishin**[2]

[1] Department of Physics, Petrozavodsk State University, Lenin Street 33, 185910, Petrozavodsk, Russia
[2] Department of Condensed Matter Physics, Royal Institute of Technology, SE-164 40 Stockholm-Kista, Sweden

*Corresponding author: e-mail alexiii1982@mail.ru, Phone: +7(8142)719681, Fax: +7(8142)711000
** e-mail velichko@psu.karelia.ru, Phone: +7(8142)635773, Fax: +7(8142)711000

We present the results on excimer laser modification and patterning of amorphous vanadium pentoxide films. Wet *positive* resist-type and Ar ion-beam *negative* resist-type etching techniques were employed to develop UV-modified films. $V_2O_5$ films were found to possess sufficient resistivity compared to standard electronic materials thus to be promising masks for sub-micron lithography.

**1 Introduction** Nowadays advances in synthesis and characterization of transition metal oxide semiconductors enable appearance of a new field of *oxide electronics* [1]. Vanadium pentoxide is one of these promising materials for microelectronic, electrochemical and optoelectronic devices [2-8]. Some promising applications of $V_2O_5$ have been discussed in our previous paper [9]. Furthermore, vanadium pentoxide can be reduced to lower oxides [2], such as, e.g., $VO_2$, and vanadium dioxide has various potential applications in electronics due to the metal-insulator transition [10]. As was noted in [9], development of lithography methods for $V_2O_5$ remains a topical problem, and in this paper we evolve and extend the results obtained there.

In principle, inorganic resists should have higher limiting resolution than polymer resists due to smaller fundamental structural units and stronger bonds in the former. [11, 12]. However, compositional and/or structural inhomogeneities may limit their ultimate resolution, while organic hybrid resists with incorporated inorganic particles can result in sub-100 nm resolution [13]. Other merit of inorganic resists is their high plasma-stability as compared to an organic resist.

Regarding the activation mechanisms, inorganic resists can be generally categorized into two groups: inorganic photo- and thermoresists. AgI, chalcogenides $GeSe_y$ and $As_2S_3$ present the group of inorganic *photoresists*. With the addition of silver, Ag-chalcogenide inorganic resists (e.g. $Ag_2Te/As_2S_3$ and $Ag_2Se/GeSe_2$) possess very high exposure sensitivity: 3 – 10 $mJ/cm^2$ at 248 nm [14]. Instead of being photochemically sensitive to light, inorganic *thermoresists* are activated by heat from a light, by electrons (*e*-beam), ions (ion-beam) or X-ray absorbed by the resist film. Since the conversion process requires simply a temperature change, thermoresists appear to be more wavelength invariant than organic or inorganic photoresists.

Group of inorganic thermoresists is presented by Fe/O, Al/O, and bimetallic resists as Bi/In and Sn/In.

A resist made of a material, which can be used for fabrication of opto-electronic devices, is of special interest [15-17]. One of such materials is just the amorphous $V_2O_5$ studied in this work.

Due to the unfilled *d*-shells, transition metals exhibit multiple oxidation states and form, as a rule, a number of oxides which can undergo various structural and phase transformations under the influence of different external actions, such as heat treatment, electron and ion bombardment, laser radiation, etc. Amorphous vanadium pentoxide has also been reported to undergo such transformations known as thermo- [5] photo- [4, 5, 20-22], and electrochromism [23, 24], electron- and ion-beam-induced modification [18, 19], and AFM patterning [25]. Recently, we have shown the possibility to perform laser UV (405 nm wavelength) micro-patterning of amorphous vanadium pentoxide films [9], 100 nm-scale resolution patterning of amorphous $V_2O_5$ and $VO_2$, prepared by anodic oxidation, using laser and *e*-beam modification [26], and memory switching with the N-type negative differential resistance, associated with the hydrogen ionic transport, in "V/hydrated anodic vanadium oxide/Au" sandwich structures [27].

Amorphous $V_2O_5$ films obtained by anodic oxidation are the most sensitive to the external perturbations [26], but the main drawback of the anodic films is the residuary metallic sublayer. Therefore, amorphous $V_2O_5$ films obtained by vacuum deposition seem to be more perspective for the processes of selective etching, micro- and nanostructure fabrication [9]. In addition, high plasma resistivity of oxides allows usage of $V_2O_5$ as an inorganic resistive mask for plasma etching.

Most previous researches were focused on electrochromic and photochromic coloration of vanadium pentoxide [4, 5, 20-24], in which transparent amorphous or crystalline thin films can be colored reversibly to a blue state. This effect is usually described in terms of the double-charge injection model. In the photochromism, water is decomposed into H and highly reactive O atoms upon irradiation of UV light and $H_xV_2O_5$ forms. (если эти рассуждения нужны, то можно оставить как есть – пусть небольшой повтор из 9 останется – ничего страшного) Present or adsorbed water plays an important role. The enhancement of photochromism in $V_2O_5$ thin films by adsorption of dimethylformamide (DMFA) molecules on the oxide film surface was reported in [28]. Short (20 ns) excimer laser pulses create vacant oxygen sites thus result in thermochromic modification of $V_2O_5$ surface [20].

Unlike our previous work [9], where the modification under the action of the low power (~20mW/cm2) UV radiation has been studied, in this work we study the effect of the intence UV excimer laser irradiation, which has a different physical nature of the influence. The powerful radiation is supposed to produce stronger structure modification all over the film thickness, higher selectivity of chemical etching and the surface morphology change. In addition, in this paper we present the results of ion beam etching and the etching rate measurements.

**2 Experimental**

$V_2O_5$ films of 100 to 300 nm thick were fabricated by pulsed laser deposition (PLD) from stoichiometric $V_2O_5$ target at room temperature onto glass, Si and $SiO_2$/Si substrates, as described in [9]. KrF ($\lambda$=248 nm) excimer laser (Lambda Physik 300) was used with a pulse duration of 20 ns. The vacuum system ensured the residual gas pressure lower than $10^{-6}$ Torr. The laser radiation energy and pulse repetition frequency were 200 mJ and 10 Hz, respectively. Laser energy density on the surface of the $V_2O_5$ target was approximately 2-4 J·cm$^{-2}$. The distance between the target and substrate was 95 mm, and the oxygen partial pressure during the deposition was 70 mTorr.

All the films prepared at room temperature have a good adhesion and withstand standard test with a scotch tape.

UV-laser modification was performed by KrF excimer laser (COMPEX-102), $\lambda$ = 248 nm, by two pulses (20 ns duration) with a total dose of 150 mJ·cm$^{-2}$. Laser spot size was about 200 mm$^2$. Exposure was performed through a special network mask with a step of 11 μm (Fig. 1).

Ion-beam etching (IBE) was employed using a standard Kaufman type ion source (Veeco Microetch) with argon pressure in the discharge and sample chambers of $10^{-4}$ and $2·10^{-6}$ Torr, respectively. Etching depth and films thicknesses were measured by a profilometer (Tencor P10).

Scanning electron microscopy (SEM) and atomic force microscopy (AFM) images were obtained using Zeiss DSM-942 and SMM-2000 microscopes, respectively.

**3. Results and discussion** AFM and SEM studies showed that the as-fabricated $V_2O_5$ films were homogeneous without pin-holes and foreign inclusions. The surface roughness before UV modification was ~ 3 nm (Fig.1, a, b), which was more than six times lower than that for the modified oxide (~ 20 nm). X-ray structural analysis showed that the before and after UV modification $V_2O_5$ films were amorphous – their X-ray $\theta$–$2\theta$ scans did not contain any diffraction peaks. The bandgap of 2.3 eV measured from the optical absorption edge corresponds to vanadium pentoxide [9].

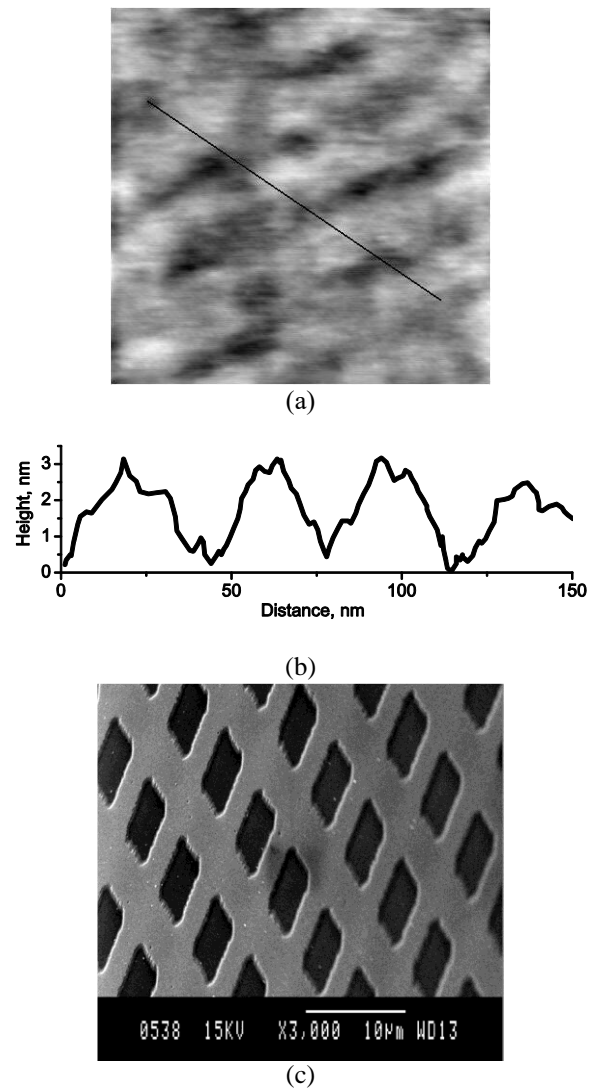

**Figure 1** a) AFM image (150×150 nm) of $V_2O_5$ films before UV modification; b) Surface profile along the line shown in Fig. 1, a; c) SEM image of a $V_2O_5$ network with a step of 11 μm after developing in formic acid-acetone-formalin solution.

Recently [30], we have shown that UV laser modification of an initial vanadium pentoxide film lead to a decrease of the short order range sizes from 2.4 nm to 1.2 nm. A minimum structure unit of the studied amorphous oxide films is a strongly distorted oxygen octahedron. The character of tetragonal pyramid distortions in initial and modified films is different, and it also differs from the $V_2O_5$ crystalline phase. Also, the small part of pyramids in the studied films of amorphous vanadium pentoxide is not completed (there is no one of tops).

Development of the modified films was accomplished in the formic acid-acetone-formalin (15:10:1) solution. This process is similar to the development of a positive resist, the virgin-to-modified selectivity of etching was found to be ~10. Figure 1 demonstrates the SEM image of the surface of $V_2O_5$ film subjected to 2 pulses of 248 nm excimer laser with the total dose of 150 mJ·cm$^{-2}$ and selectively etched in the formic acid-acetone-formalin solution.

Amorphous $V_2O_5$ resist can be easily removed in, e.g., 1% aqueous HCl, and crystallized $V_2O_5$ resist – in the solution NaOH-$H_2O_2$-$H_2O$ (1:1:100); note that similar compositions are usually used for developing of standard photoresists without damage of Si substrates.

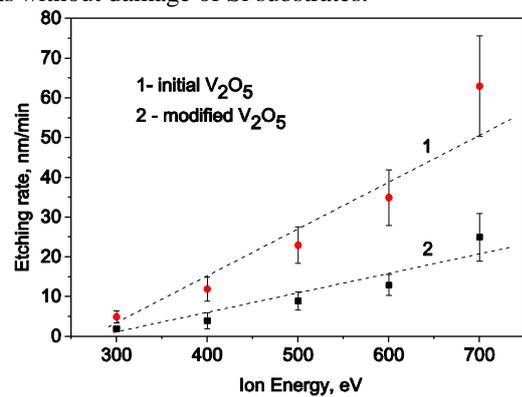

(a)

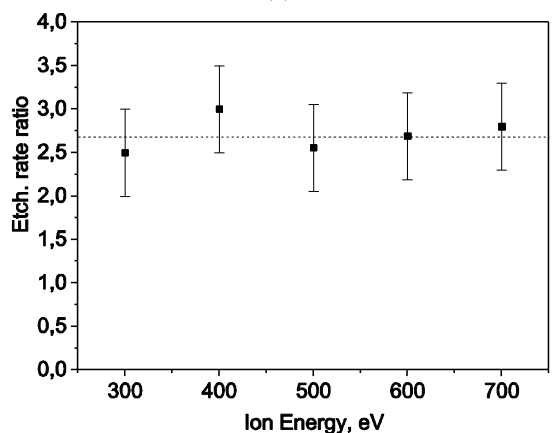

(b)

**Figure 2, a** - Etching rate of initial and modified $V_2O_5$; **b** - IBE selectivity as a function of energy of Ar ion beam.

Special attention was paid to study the resistivity of UV modified $V_2O_5$ against Ar ion-beam etching (IBE). Figure 2 (a) demonstrates significant growth of the etching rate both in virgin and modified $V_2O_5$ films with increasing of Ar ion beam energy (300 – 700 eV). This process is similar to the development of a negative resist, the modified-to-virgin selectivity of $V_2O_5$ films etching was found to be 2.5−3.0 (see Fig.2, b). The selectivity achieved in IBE developing of vanadium pentoxide resist appears to be lower than that for the wet etching, but in some cases it can be used with sufficient efficiency.

Table 1 demonstrates the rates of Ar-IBE for exposed, initial $V_2O_5$ (authors data) and some other materials [31] at ion-beam power density w = 0.5 W/cm$^2$, vertical incidence of ions with energy 450 eV, and residual pressure in a chamber less than 10$^{-4}$ Pa.

It becomes evident that the rate of etching for the exposed $V_2O_5$ oxide is 3 - 4 times lower than that for most of the materials applied in microelectronic industry. Especially high selectivity appears in comparison with such semiconductors as GaAs, InAs, and InSb widely applied in optoelectronic devices [32].

**Table 1** Comparison of Ar ion-beam $V_2O_5$ etching rates with etching rates of various electronic materials [31].

| Material | Etching rate [nm/min] | Selectivity of material in relation to UV-laser exposed $V_2O_5$ |
|---|---|---|
| exposed $V_2O_5$ | 6-7 | - |
| initial $V_2O_5$ | 17-19 | 2.5-3.0 |
| GaAs | 132-150 | 20-23 |
| Ag | 102-150 | 15-23 |
| LiNbO$_3$ | 30 | 4.6 |
| Si | 18-30 | 2.8-4.6 |
| Ti | 18 | 2.8 |
| SiO$_2$ (thermal) | 24-36 | 3.7-5.5 |
| Permalloy | 18-30 | 2.8-4.6 |
| V | 9-18 | 1.4-2.8 |
| PMMA | 42-48 | 6.5-7.4 |

Also, one can see from Table 1 that $V_2O_5$ possesses higher resistivity to ion-beam etching (6-7 times) as compared to PMMA.

### 4. Conclusions

In summary, we demonstrate the possibility to use $V_2O_5$ films as UV resist. Modification of amorphous layers of $V_2O_5$ was made by 248 nm excimer laser source. Two methods of development of UV modified $V_2O_5$ films were employed: *positive* resist-type wet etching in the formic acid-acetone-formalin solution and *negative* resist-type Ar ion-beam etching. Sufficient resistivity of amorphous $V_2O_5$ films compared to standard electronic materials makes them perspective for sub-micron fabrication.

**Acknowledgements** This work was supported by Svenska Institutet (Dnr: 01370/2006), the Ministry of Education and Science of Russian Federation through the "Development of Scientific Potential of High School" Program (projects No. 4978 and 8051) and the U.S. CRDF grant (award No. RUX0-013-PZ-06). Also, the authors thank Dr. B. Loginov and Dr. S. Bonetti for their help with AFM and SEM experiments.

**References**

[1] *Ramirez. A. P.* Oxide Electronics Emerge // Science. 2007. V. 15. P. 1377.
[2] A. Z. Moshfegh and A. Ignatiev, Thin Solid Films **198**, 251 (1991).
[3] M. Benmoussa, A. Outzourhit, A. Bennouna, and E. L. Ameziane, Thin Solid Films **405**, 11 (2002).
[4] A. I. Gavrilyuk, A. A. Mansurov, and F. A. Chudnovskii, Sov. Tech. Phys. Lett. **10**, 292 (1984).
[5] A. I. Gavrilyuk, N. M. Reinov, and F. A. Chudnovskii, Sov. Tech. Phys. Lett. **5**, 514 (1979).
[6] Yoshitaka Fujita, Katsuhiro Miyazaki, and Chiei Tatsuyama, Japan J. Appl. Phys. Part 1 **24**, 1082 (1985).
[7] C. Julien, E. Haro-Poniatowski, M. A. Camacho-Lopez, L. Escobar-Alarcon, and J. Jimenez-Jarquin, Materials Science and Engineering **B65**, 170 (1999).
[8] Han Young Yu, Byung Hyun Kang, Ung Hwan Pi, Chan Woo Park, Sung-Yool Choi, and Gyu Tae Kim, Appl. Phys. Lett. **86**, 253102 (2005).
[9] V. V. Putrolaynen, A. A. Velichko, A. L. Pergament, A. B. Cheremisin, and A. M. Grishin, J. Phys. D: Appl.Phys. **40**, 5283 (2007).
[10] M. Imada, A. Fujimori and, Y. Tokura, Rev. Mod. Phys. **70,** 1059 (1998).
[11] H. Jain, M. Vlcek, J. of Non-Cryst. Sol. **354**, 1401 (2008).
[12] K. Ogino, J. Taniguchi, S. Satake, K. Yamamoto, Y. Ishii, K. Ishikawa, Microelectronic Engineering **84,** 1071 (2007).
[13] L. Merharia, K.E. Gonsalvesb, Y. Huc, W. Hec, W.-S. Huangd, M. Angelopoulose, W.H. Bruenger, C. Dzionk, M. Torkler, Microelectronic Engineering **63**, 391 (2002).
[14] J. M. Lavine and M. J. Buliszak, J. Vac. Sci. & Technol. B **14**, 3489 (1996).
[15] R.Houbertz, G.Domanna, C.Cr onauera, A.Schmitt, H.Martin, J.-U. Parkb, L.Frohlich, R.Buestricha, M.Popalla, U.Streppel, P.Dannber, C.Wachter, A.Brauer, Thin Sol. Films **442,** 194 (2003).
[16] G. Brusatin, G. Della Giustina, F. Romanato, M. Guglielmi, Nanotechnology, **19,** 175306 (2008).
[17] G. Della Giustina, G. Brusatin, M. Guglielmi and F. Romanato, Materials Science and Engineering: C **27,** 1382 (2007).
[18] K. Nobuyoshi, O. Koichi, A. Masanobu, K. Masanori, and A. Nobufumi, Japan. J. Appl. Phys. **27**, 314 (1988).
[19] D. J. Smith, M. R. McCartney, and L. A. Bursill, Ultramicroscopy **23**, 299 (1987).
[20] Z. Liu, G. Fang, Y. Wang, Y. Bai, and K.-L. Yao, J. Phys. D: Appl. Phys. **33**, 2327 (2000).
[21] [21] F. A. Chudnovskii, A. L. Pergament, D. A. Schaefer, and G. B. Stefanovich, J. Sol. St. Chem. **118**, 417 (1995).
[22] S. Nishio and M. Kakihana, Chem. Mater. **14**, 3730 (2002).
[23] K. Nagase, Y. Shimizu, N. Miura, and N. Yamazoe, Appl. Phys. Lett. **60,** 802 (1992).
[24] Q. H. Wu, A. Thiβen, and W. Jaegermann, Sol. St. Ion. **167**, 155 (2004).
[25] S. Iwanaga, R. B. Darling, and D. H. Cobden, Appl. Phys. Lett. **86**, 133113 (2005).
[26] G. B. Stefanovich, A. L. Pergament, A. A. Velichko, and L. A. Stefanovich, J. Phys.: Cond. Matt. **16**, 4013 (2004).
[27] A. Velichko, V. Putrolaynen, G. Stefanovich, N. Kuldin, A. Cheremisin, I. Feklistov and N. Khomlyuk, J. Phys. D: Appl. Phys. **41**, 225306 (2008).
[28] A. Gavrilyuk, SPIE **2968**, 195 (1997).
[29] S. A. Aly, S. A. Mahmoud, N. Z. El-Sayed, and M. A. Kaid, Vacuum **55**, 159 (1999).
[30] A. B. Cheremisin, S. V. Loginova, A. A. Velichko, V. V. Putrolaynen, A. L. Pergament and A. M. Grishin, J. Phys.: Conf. Ser. **100**, 052096 (2008).
[31] B. S. Danilin and V. J. Kireev, Application of low temperature plasma for etching and cleaning of materials (Energoatomizdat, Moscow, 1987).
[32] F. Frost, A. Schindler, and F. Bigl, Semicond. Sci. Technol. **13**, 523 (1998).